\documentclass[cits]{PoS}

\newcommand\url{}%
\newcommand\arcdeg{\mbox{$^\circ$}}%
\newcommand{\uJy}{\ensuremath{\mu{\rm Jy}}}

\title{The ATLAS Survey of the CDFS and ELAIS-S1 Fields}

\ShortTitle{ATLAS (Australia Telescope Large Area Survey)}

\author{\speaker{Emil Lenc}, Ray Norris\\
        Australia Telescope National Facility\\
        E-mail: \email{Emil.Lenc@csiro.au}, \email{Ray.Norris@csiro.au}}

\author{Chris Hales, Kate Randall\\
        University of Sydney\\
        E-mail: \email{chales@physics.usyd.edu.au}, \email{krandall@physics.usyd.edu.au}}

\author{Andrew Hopkins, Rob Sharp\\
        Anglo-Australian Observatory\\
        E-mail: \email{ahopkins@aao.gov.au}, \email{rgs@aao.gov.au}}

\author{Minh Huynh\\
        Infrared Processing and Analysis Center\\
        E-mail: \email{mhuynh@ipac.caltech.edu}}

\author{Minnie Mao\\
        University of Tasmania\\
        E-mail: \email{mymao@utas.edu.au}}

\author{Enno Middelberg\\
        Astronomisches Institut, Ruhr-Universit\"{a}t Bochum\\
        E-mail: \email{middelberg@astro.rub.de}}

\abstract{The first phase of the ATLAS (Australia Telescope Large Area Survey) project surveyed a total 7 square degrees down to 30 $\uJy$ rms at 1.4 GHz and is the largest sensitive radio survey ever attempted. We report on the scientific achievements of ATLAS to date and plans to extend the project as a path finder for the proposed EMU (Evolutionary map of the Universe) project which has been designed to use ASKAP (Australian Square Kilometre Array Pathfinder).}

\FullConference{Panoramic Radio Astronomy: Wide-field 1-2 GHz research on galaxy evolution\\
		 June 2-5 2009\\
		 Groningen, the Netherlands}

\begin{document}

\section{Introduction}

The first phase of the ATLAS (Australia Telescope Large Area Survey) project has surveyed a total 7 square degrees down to 30 $\uJy$ rms at 1.4 GHz and is the largest sensitive radio survey ever attempted. The survey has observed an area surrounding the CDFS (Chandra Deep Field South) and ELAIS-S1 (European Large Area ISO Survey - South 1) regions with the ATCA (Australia Telescope Compact Array). The survey areas were chosen to cover the Southern SWIRE fields, which have deep optical, near-infrared, and far-infrared (and in some parts of the field, deep X-ray) data, so that this combined SWIRE/ATLAS survey may be the most comprehensive multi-wavelength survey yet attempted.

\begin{figure}[ht]
\includegraphics[width=1.0\textwidth]{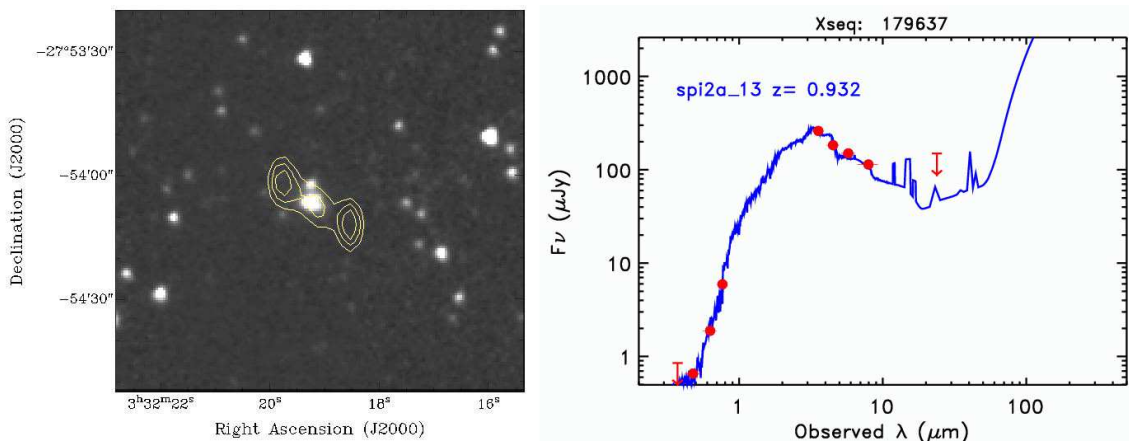}
\caption{ATLAS contains unusual sources, such as these Powerful Radio Objects Nested in Galaxies with Star formation (PRONGS). These sources have the radio appearance of an AGN (left), but have a spectral energy density typical of classical star-forming galaxies (right), with no hint of an AGN \cite{nor08}.}
\label{fig:fig1}
\end{figure}

The broad scientific goal of ATLAS is to understand the evolution of galaxies in the early Universe. The radio observations are important as they penetrate the heavy dust extinction found in the most active galaxies at all redshifts \cite{cha03} and are particularly effective at detecting AGN buried within dusty galaxies (Figure \ref{fig:fig1}).

\section{Outcomes}
The ATLAS team have already achieved the following:
\begin{itemize}
\item The first CDFS and ELAIS-S1 survey results have been published \cite{nor06,mid08a}.
\item A new and unexpected class of object, Infrared-Faint Radio Sources (IFRS), have been discovered \cite{nor06}, catalogued and analysed \cite{huy09}, and observed using VLBI \cite{nor07,mid08b}.
\item Discovered and explained a change in slope of the FIR-radio correlation at $S_{1.4}<1$ mJy \cite{nor09,boy07}.
\item Discovered a z $\sim0.2$ cluster through the presence of a wide-angled tail ATLAS radio source \cite{mao10}.
\item Begun analyses on the evolution of radio sources \cite{mao09}, and to optimise discrimination between AGN and star forming galaxies \cite{ran09}.
\item Produced images and catalogues of source polarisation which we will use to investigate individual sources. By using the rotation measure grid we hope to measure or put limits on the intergalactic magnetic field \cite{hal10}.
\end{itemize}

\section{Current Status}
The CDFS and ELAIS-S1 fields have been imaged down to 30 $\uJy$ rms at 1.4 GHz yielding a total of $\sim2000$ radio sources. Catalogues of these sources have been made publicly available through NED (the NASA/IPAC Extragalactic Database) and work is currently underway to provide VO (Virtual Observatory) access to the catalogues also.

\begin{figure}[ht]
\includegraphics[width=0.51\textwidth]{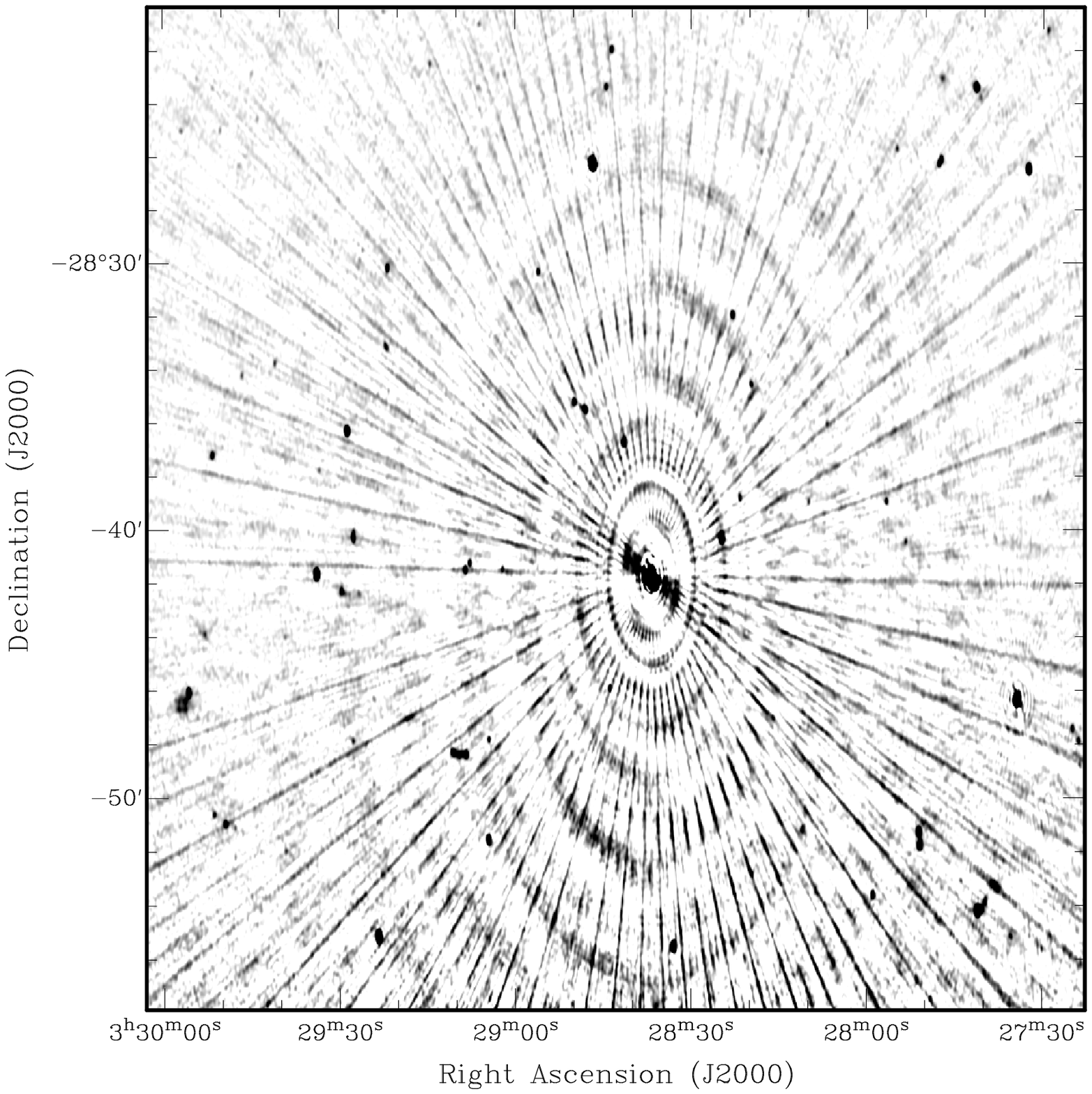}
\includegraphics[width=0.51\textwidth]{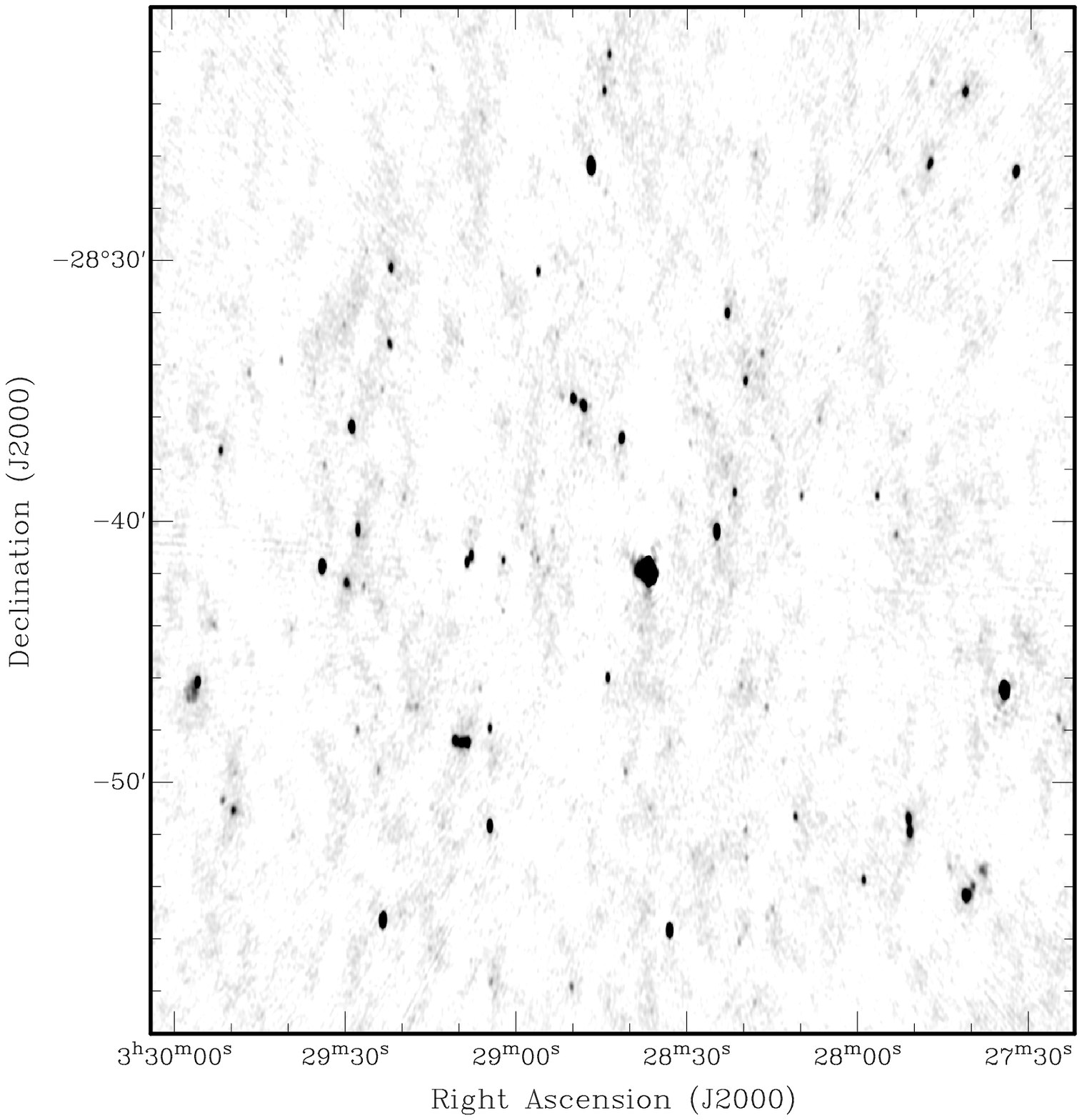}
\caption{(a) Left, image of confusing source from first ATLAS data release. (b) Right, image with improved source modelling applied.}
\label{fig:fig2}
\end{figure}

The 30 $\uJy$ rms limit of the first ATLAS data release was generally imposed by observational sensitivity, however, it was clear that further observations would be limited by our ability to mitigate the effects of bright sources in each field (Figure \ref{fig:fig2}a). We have recently been successful in removing these limitations in one pointing of the CDFS region by modelling bright confusing sources in the u-v plane. This enabled self-calibration to converge on a more optimal solution, and after subsequent deconvolution, allowed the theoretical signal-to-noise to be achieved throughout the entire field (Figure \ref{fig:fig2}b). We anticipate releasing an updated catalogue, utilising this improved imaging, by 2010.

We are now in the process of continuing observations of both ATLAS fields to achieve the originally proposed goal of 10 $\uJy$ rms over the full 7 sq. deg. These observations will make use of the Compact Array Broadband Back-end (CABB) which provides a continuous 0.5 GHz band at 1.4 GHz. We estimate the detection of ~10000 radio sources with these observations.

\clearpage
\section{ATLAS as an ASKAP Path-finder}
EMU (Evolutionary Map of the Universe) is a proposed radio sky survey project designed to use the new ASKAP (Australian Square Kilometre Array Pathfinder) telescope. The EMU-wide project, will make a deep ($\sim10$ $\uJy$ rms) radio continuum survey covering the entire Southern Sky as far North as $30\arcdeg$. As a result, it will be able to probe star forming galaxies up to z$=1$, AGNs to the edge of the Universe, and will undoubtedly uncover new classes of object. Similarly, the EMU-deep project will aim to observe a smaller region of the sky ($\sim30$ sq. degrees) down to an rms of about $\sim1$ $\uJy$.

It is envisaged that by pushing the ATLAS fields down to 10 $\uJy$ rms we will be using it as a design study for ASKAP EMU-wide. This will result in a better characterisation of the challenges faced by deep surveys, such as EMU-wide and EMU-deep e.g. the effects of confusion. Subsequently, ATLAS will drive the development of new advanced calibration and imaging algorithms that will ultimately be used for future instruments such as ASKAP. The ATLAS project also plans to set up a pipeline for the automatic multi-wavelength cross-identification of radio sources. The output of this pipeline will be placed in the public domain in the form of images and source catalogues. Finally, ATLAS will provide the first preview of the science that can be achieved with surveys such as EMU. So watch this space, ATLAS has a great deal to offer over the next few years.

\end{document}